\documentclass[conference]{IEEEtran}
\IEEEoverridecommandlockouts
\usepackage{tabularx}
\usepackage{multicol}
\usepackage{multirow}
\usepackage{cite}
\usepackage{amsmath,amssymb,amsfonts}
\usepackage{hyperref}
\usepackage{url}
\usepackage{balance}
\usepackage{algorithmic}
\usepackage{graphicx}
\usepackage{textcomp}
\usepackage{xcolor}

\def\BibTeX{{\rm B\kern-.05em{\sc i\kern-.025em b}\kern-.08em
    T\kern-.1667em\lower.7ex\hbox{E}\kern-.125emX}}
\begin{document}

\title{Direct-to-Device Connectivity for Integrated Communication, Navigation and Surveillance\\
}
\author{\IEEEauthorblockN{
Muhammad Asad Ullah\IEEEauthorrefmark{1}, Davi Brilhante\IEEEauthorrefmark{2}, Luís Eduardo Partichelli Potrich\IEEEauthorrefmark{2}\IEEEauthorrefmark{4},\\José Suárez-Varela\IEEEauthorrefmark{3}, Paul Almasan\IEEEauthorrefmark{3}, Charles Cleary\IEEEauthorrefmark{2}, Vadim Kramar\IEEEauthorrefmark{1}\\
}
\IEEEauthorblockA{\IEEEauthorrefmark{1}VTT Technical Research Centre of Finland, Finland}
\IEEEauthorblockA{\IEEEauthorrefmark{2}ART, Collins Aerospace, Ireland}
\IEEEauthorblockA{\IEEEauthorrefmark{3}Telefonica Research, Spain}
% John Dooley\IEEEauthorrefmark{4},
\IEEEauthorblockA{\IEEEauthorrefmark{4}Maynooth University, Ireland}
}

%\author{\IEEEauthorblockN{Muhammad Asad Ullah, Vadim Kramar}
%\IEEEauthorblockA{
%\textit{VTT Technical Research Centre of Finland}\\
%Espoo, Finland\\}
%\and
%\IEEEauthorblockN{José Suárez-Varela, Paul Almasan}
%\IEEEauthorblockA{
%\textit{Telefonica Research}\\
%Madrid, Spain}
%\and
%\IEEEauthorblockN{Davi Brilhante, Charles Cleary, \\Luís Eduardo Partichelli Potrich}
%\IEEEauthorblockA{
%\textit{ART, Collins Aerospace}\\
%Cork, Ireland}

%}

\maketitle

\begin{abstract}
Sixth-generation (6G) communication systems are expected to support direct-to-device (D2D) connectivity, enabling standard user equipment (UE) to seamlessly transition to non-terrestrial network (NTN), particularly satellite communication mode, when operating beyond terrestrial network (TN) coverage. This D2D concept does not require hardware modifications to conventional UEs and eliminates the need for dedicated satellite ground terminals. D2D-capable UEs can be mounted on both manned and unmanned aircraft, however, they are especially well-suited for low-altitude unmanned aircraft due to their compact form factor, lightweight design, energy efficiency, and TN–NTN roaming capabilities. D2D can also enable beyond-visual-line-of-sight operation by providing NTN support for Communications, Navigation, and Surveillance (CNS) services during TN outages or congestion. This paper investigates the capabilities and limitations of D2D connectivity for low-altitude unmanned aircraft operating in urban environments. We analyze the variation in line-of-sight probability for both TN and NTN links as a function of aircraft altitude. We further compute path loss and received signal strength while accounting for a representative TN deployment with down-tilted antennas. 
The results show that the TN and NTN links complement each other, significantly improving the availability of the CNS service at low altitudes. These findings provide insights to support the design and optimization of future 6G-enabled integrated CNS services.
\end{abstract}

\section{Introduction}

Modern aircraft operations do not fully benefit from the use of 3rd generation partnership project (3GPP) technologies despite rapid advancements in the standards. The coverage of the cellular terrestrial network (TN) is optimized to serve users located at ground level and in the buildings, with down-tilted antennas \cite{Giovanni,Chen}. Non-terrestrial networks (NTN) address this challenge by extending wireless coverage to ground and aerial users, including the altitudes at which unmanned and manned aircraft operate~\cite{Leon,figaro2026}. %at which unmanned and manned aircraft operate~\cite{figaro2026}. %It is being discussed that NTN will be a native and an integral component of the 6G networks, where every smart device will support both TN and NTN based cellular connections. 
Moreover, NTNs are increasingly viewed as a native component of 6G networks, in which smart devices are expected to support both TN- and NTN-based cellular connectivity. This means 6G will benefit from the resources of both network segments to ensure service
availability, continuity, ubiquity, and scalability~\cite{Araniti}.
\begingroup
\renewcommand{\thefootnote}{}
\footnotetext{This work has been accepted in IEEE 26th Integrated Communications, Navigation and Surveillance Conference, April 14--16, 2026, Herndon, Virginia, USA. Copyright has been transferred to IEEE.}
\endgroup
Direct-to-device (D2D) connectivity is an emerging concept in 6G, where a standard consumer device, also called user equipment (UE), connects directly to a satellite when outside the TN coverage without needing any terrestrial base station~(TBS) or ground satellite terminal~\cite{pasandi2024survey,ullah3gpp}. A D2D-capable UE can seamlessly switch between TN and NTN by leveraging its roaming capabilities~\cite{ESA_D2D_2026}, making it a promising approach to serve Unmanned Aircraft System~(UAS) Traffic Management (UTM) with Communications, Navigation, and Surveillance (CNS) services.\footnote{Hereafter, we use the term aircraft to refer to those that operated at low-altitudes, including unmanned aircraft systems (UAS), vertical take-off and landing (VTOL)-capable aircraft (VCA), and general aviation, such as helicopters.} This paves the way for implementing the Integrated CNS (ICNS) concept, which aims to deliver the ‘C’, ‘N’, and ‘S’ services jointly through the same technology stack~\cite{AsadCNS,Alshaer}. In addition to a wide range of applications, D2D-capable UEs will be able to establish TN and NTN links through the 3GPP stack to deliver ICNS services.

In the communications domain, UAS require low latency, high reliability, and continuous connectivity to support command-and-control (C2) operations and meet application-level data transmission requirements. %Service continuity is a fundamental requirement of the ICNS system. 
By leveraging the complementary resources of TN and NTN, a UE with D2D functionalities can play a key role in ensuring uninterrupted ICNS service. For example, a UE mounted on an aircraft is expected to rely primarily on TN infrastructure to deliver CNS data. In the absence of terrestrial coverage, the UE can seamlessly switch to the NTN system, relying on low Earth orbit (LEO) satellites to maintain connectivity.

For navigation, stringent positioning accuracy requirements are currently under discussion for 6G systems. According to 3GPP TR 38.914 \cite{3gppTR38914}, the target positioning accuracy is 50 $\mathrm{meters}$ ($\mathrm{m}$) for manned aircraft en-route operations, 10 $\mathrm{m}$ for its landing, and as low as 1 $\mathrm{m}$ for unmanned aircraft operations. 3GPP TR 22.870\cite{3gppTR22870} discusses accurate and latency positioning services for aircraft using hybrid TN–NTN technologies. A D2D-capable 6G UE can take advantage of hybrid TN–NTN connectivity to enable highly accurate positioning services, thus meeting the strict requirements of unmanned aircraft navigation and surveillance. %Nevertheless, further research is required to thoroughly validate this idea. %positioning for aircraft targets an accuracy of 3 $\mathrm{m}$ with 99\% service availability and a service latency between 0.1 and 0.5 seconds, while supporting aircraft speeds of up to 160 $\mathrm{km/h}$. Nevertheless, further research is required to comprehensively %evaluate and validate these key performance indicators.

In terms of surveillance, a 6G UE equipped with D2D capabilities can enhance situational awareness in digital airspace by ensuring continuous and reliable connectivity, which is a key requirement for real-time aircraft monitoring and tracking~\cite{AsadCNS}. 

The main contributions of this paper are as follows:
\begin{itemize}
    \item We present an ICNS vision in which a UE mounted on an aircraft is equipped with D2D capabilities and can seamlessly roam between TN and NTN systems, thereby enhancing coverage availability and reliability by benefiting from network resources of both segments.
    \item We propose and develop a statistical framework to analyze and characterize the connectivity limits of TN and NTN systems while delivering ICNS services to an aircraft operating at low altitude in an urban environment.% Second, we develop a statistical framework to investigate and identify the limits of TN and NTN connection for delivering ICNS services to a aircraft operating at low altitude in urban environment.}
    \item We validate the proposed framework over scenarios derived from relevant 3GPP technical reports and specifications. Particularly, we adopt 3GPP TR~38.811 \cite{3gppTR38811}, TR~36.814\cite{3gppTR36814}, and TS~38.108\cite{3gppTS38108} to model path loss, vertical antenna gain, and receiver sensitivity thresholds, respectively. We then assess coverage by comparing the received signal strength indicator (RSSI) against sensitivity thresholds for both TN and NTN systems.
\end{itemize}

Overall, our results indicate that UEs with D2D capability and seamless roaming between TN and NTN provide a feasible solution for delivering ICNS services at low altitudes.

The remainder of this paper is organized as follows. In Section~\ref{sec:sec2} we present the technical background and related works. In Sections~\ref{sec:sec3} and \ref{sec:sec4}, we describe our system model and discuss the results. Section~\ref{sec:sec5} presents a discussion of the findings and highlights the key lessons learned. Finally, Section~\ref{sec:sec6} concludes this paper with final remarks.

\section{Technical Background and Related work}
\label{sec:sec2}

This section discusses the studies on UAS and NTN channel modeling and recent developments of hybrid TN and NTN for the connectivity needs of UAS. Table~\ref{table:tab1} summarizes the key aviation acronyms and definitions used throughout the paper.

\begin{table}[t!]
    \centering
    \caption{The list of key Aviation Acronyms and Definitions.}
    \resizebox{\columnwidth}{!}{\begin{tabular}{|l|l|}
       \hline
       \textbf{Acronyms} & \textbf{Definition} \\ \hline
       3GPP  & 3rd generation partnership project \\
       5G    & Fifth generation of cellular network technology \\
       6G    & Sixth  generation of cellular network technology \\
       A2A   & Aircraft-to-aircraft\\
       A2X   &  Aircraft-to-everything\\
       A2G   & Air-to-ground\\
       ADS-B & Automatic dependent surveillance-broadcast \\
       ACARS & Aircraft communications addressing and reporting system\\
       %ATM   & Air Traffic Management \\
       C2    & Command and control link\\
       CDL   & Clustered delay line\\
       CNS   & Communication, navigation and surveillance \\
       CNPC  & Command and non-payload communication\\
       D2D   & Direct-to-device\\
       GNSS  & Global navigation satellite system \\ 
       ICNS  & Integrated CNS\\
       LoS   & Line-of-sight \\
       mmWave & millimeter wave \\
       NTN   & Non-terrestrial network\\
       TBS   & Terrestrial base stations\\
       TDL   & Tapped delay line \\
       U-space & U-space airspace and digital services \\
       UAS   & Unmanned aircraft system\\
       UTM   & UAS traffic management\\
       VCA   & Vertical take-off and landing (VTOL) capable aircraft\\
       VLL   & Very low-level\\
        \hline
    \end{tabular}}
    \label{table:tab1}
\end{table}

\subsection{UAS and NTN channel modeling}
\label{sec:sec2a}
3GPP has conducted investigations and specification works to enable connectivity for UAS, initially in LTE and later in 5G New Radio (NR). Early efforts in Release 15 focused on enhanced LTE support for aerial vehicles, captured in TR 36.777~\cite{3gppTR36777}, which defines evaluation scenarios, height-dependent propagation and line-of-sight (LoS) models, and identifies key interference and mobility challenges when serving unmanned aircraft with terrestrial base stations (TBSs) optimized for ground users. This report presents a channel model for user heights above the nominal ground level (typically 1.5~$\mathrm{m}$), based on a UAS measurement campaign conducted across multiple field-trial configurations that vary the scenario, altitude, and mobility. Extensive results are provided in terms of reference signal received power~(RSRP), reference signal received quality~(RSRQ), and received signal strength indicator~(RSSI). %In addition, simulations were performed to provide calibration results for %system-level simulations using 
%the channel model specified in this technical report.

Likewise, TR 38.811 \cite{3gppTR38811} investigates channel and system aspects for NTN. For link-level simulations, it adopts reference Tapped delay line (TDL)/Clustered delay line (CDL) channel models and introduces elevation angle dependent LoS probabilities and path loss formulations, thus the LoS likelihood increases with elevation angle ($90^\circ$ means that the satellite is at Nadir angle) and UE environment type (from rural/suburban to dense urban). Although not focused on aerial users, this framework defines NTN system-level simulation guidelines under realistic propagation, and it has been widely used as a baseline for subsequent NTN channel modeling.
%{\color{red}unmanned aircraft height and decreases with horizontal distance. Remove.}

Beyond channel characterization, 3GPP has specified the services and architecture for UAS connectivity, identification, and tracking. In \cite{AsadCNS}, the authors list 3GPP technical reports, specifications, and related work that define mechanisms for UAS registration, remote identification, user tracking via cellular infrastructure, and integration with external UAS traffic management (UTM/U-space) systems.  A broader review of standardization efforts in \cite{Abdalla} also highlights the work of other institutions, such as IEEE and ITU. These standardization efforts indicate that current cellular systems are designed to support C2, telemetry, and identification services for UAS and low-altitude aircraft, although the focus is primarily on connectivity rather than on CNS requirements.

In parallel with standardization activities, academia has produced a rich body of work on analytical and simulation-based models for channel, propagation, and coverage analysis of unmanned aircraft communications. In~\cite{khuwaja}, authors provide a survey of unmanned aircraft channel modeling, covering analytical, geometry-based, measurement-based, and stochastic approaches for A2G and A2A links over a wide range of frequencies and environments. This survey highlights the critical role of LoS probability, altitude-dependent path loss, and three-dimensional (3D) spatial modeling in capturing the unique characteristics of aircraft channels, including dynamic low-altitude mobility. It also emphasizes the gap between terrestrial channel models and aerial-user-focused models.

LoS probability modeling has been studied both analytically and empirically. In \cite{Alhourani}, the authors derive a compact elevation angle-dependent LoS probability in urban environments; the LoS probability increases with aircraft altitude and is parameterized by building height and density statistics, capturing the notion that obstacles cause less blockage as unmanned aircraft climb to higher altitudes. In~\cite{Gapeyenko}, a study investigates a millimeter-wave~(mmWave) unmanned aircraft communications scenario in urban-grid deployments. Their approach combines stochastic geometry with 3D building layouts to characterize the LoS probability for aerial users at different heights and street locations as a function of building density. In \cite{Cui}, a 3GPP UAS-focused channel model is applied to coverage analysis by comparing an analytically derived coverage probability model with Monte Carlo simulations based on the system model presented in~\cite{3gppTR36777}. The authors evaluate the impact of key system-model components, such as antenna array size, path loss, and LoS probability, and provide design recommendations, including the use of antenna arrays with fewer elements to extend coverage due to their larger beamwidth.

These works highlight the limitations of TN coverage models, tailored to ground users, and evidence that accurately capturing unmanned aircraft mobility remains an open challenge for both industry and academia. Reliable modeling depends critically on accurate representations of the underlying network, scenario, and LoS probability model. NTN emerges as a viable solution for coverage in situations where TN cannot achieve the desired signal quality; however, the channel models in the literature primarily focus on ground users. Extending 3GPP-based models with a detailed understanding of LoS probability, path loss, and coverage for unmanned aircraft communications is a viable approach to capture more diverse environments, deployment scenarios, and UAS dynamics, motivating the specialized modeling and design considered in this paper.

\subsection{Coverage in the skies: Hybrid TN and NTN system}
\label{sec:sec2b}
Typically, TN serves as the primary connectivity solution in areas with ground infrastructure. On the other hand, NTN and, especially, satellites extend coverage to remote and oceanic areas~\cite{Araniti}. Beyond coverage extension, NTN can also complement TN in dense urban environments by managing excess traffic during peak-demand periods. Furthermore, in the event of TN failures, satellite links can maintain service continuity, thereby improving overall system resilience~\cite{Giovanni}. Hybrid TN and NTN enables the delivery of ICNS services anywhere at any time~\cite{AsadCNS}.

\subsubsection{Terrestrial network}
When TN infrastructure is present, TBS can provide an aircraft connection to deliver CNS services. In addition, TN can also enable Internet connectivity on VCA for passengers and onboard systems. It is worth mentioning that current TBS deployments are primarily optimized for ground users rather than aerial platforms~\cite{Cui,Cherif}.

\subsubsection{Non terrestrial network}
In legacy CNS, satellite communication is an important component of aeronautical communications in oceanic and remote environments. For decades, satellites have provided a communication bridge between aircraft and ground systems~\cite{Nils}. Table~\ref{table:tab2} shows the satellite technologies, constellation and their CNS services. For example, satellite-based aeronautical communication systems introduced in the early 1990s have supported voice and data services, including ACARS and ADS-B messaging. With rapid advancements in satellite technology, especially LEO constellations, NTN coverage is expanding beyond remote regions to include urban environments. This evolution enables NTN to complement TN not only in coverage-limited areas but also during network congestion or outages in urban areas~\cite{AsadCNS}.

\subsubsection{Integrated TN and NTN system}
A combination of cellular TN and proprietary NTN can be used for delivering CNS services to an unmanned aircraft flying at low altitude. However, this implementation may require two separate transceivers for TN and NTN systems~\cite{Arroyo,figaro2026}, which can be challenging for small platforms with strict size, weight, and power constraints. Existing studies highlight that combining 3GPP cellular TN and NTN systems can significantly improve service availability, reliability, and coverage continuity \cite{Araniti,rinaldi2020non}. To realize the ICNS vision, hybrid TN and NTN systems can support aircraft operations across wide geographic regions and at different altitudes, from very low-level (VLL) to high-altitude airspace~\cite{AsadCNS}.

In future communication systems, D2D capabilities will allow commercial off-the-shelf UE to transition seamlessly between terrestrial and satellite links without requiring dedicated satellite hardware \cite{ullah3gpp,Taha,Feng}. This capability significantly simplifies the use of TN and NTN, and broadens the applicability of hybrid architectures for aviation use cases, especially for delivering ICNS services to unmanned aircraft.

\begin{table}[!t]
\caption{Existing NTN technologies and their CNS services.}
\centering
  \resizebox{\columnwidth}{!}{\begin{tabular}{|l|c|c|}
\hline
\textbf{Technology} & \textbf{Constellation} & \textbf{Category}\\

\hline
\multirow{2}{*}{Classic Aero} &  \multirow{2}{*}{Viasat (Inmarsat)} & Communication\\
&  &  Surveillance\\
\hline
\multirow{2}{*}{SwiftBroadband-Safety} & \multirow{2}{*}{Viasat (Inmarsat)} &Communication\\
&& Surveillance\\
\hline
Iridium Certus &Iridium& Communication\\
\hline
Space-based ADS-B & Iridium (Aireon)&Surveillance\\
\hline
\multirow{2}{*}{GNSS} &GPS, GLONASS & Navigation\\
&
Galileo, BeiDou & 
enables Surveillance\\
\hline
\end{tabular}}
\label{table:tab2}
\end{table}
\subsection{Spectrum bands for TN and NTN}
\label{sec:sec2c}
3GPP 5G NR operates across multiple frequency ranges~(FRs) for terrestrial networks. These ranges include FR1 and FR2, as well as the upper mid-band, often referred to as FR3. FR1 spans 410\text{--}7125~$\mathrm{MHz}$ and is widely used because it offers a practical balance between coverage and capacity. FR3 fills the gap between FR1 and FR2 by operating in the 7\text{--}24~GHz band, which is better suited to more localized, high‑capacity deployments. FR2 covers 24.25\text{--}52.6~$\mathrm{GHz}$, enabling much higher data rates, but usually with shorter range and stronger sensitivity to blockage \cite{9074973}. In an urban aerial scenario with operations in low altitudes, propagation differs from that of typical ground users~\cite{11211403}. In FR1, airborne UEs often have LoS to several base stations, thereby increasing inter‑cell interference. FR2 links rely on narrower beams, which can reduce unwanted visibility and help manage interference, although coverage continuity is more challenging in dense urban areas. FR2 and FR3 links may require very-small-aperture terminals, while L/S bands in FR1 are suitable for D2D connectivity \cite{rappaport,YastrebovaCastillo2024EASN}.

NTN communication uses satellite bands specified in 3GPP TS 38.101‑5 \cite{3gppTR38101-5} and summarized in Table~\ref{tab:tab3}. The spectrum includes FR1 NTN from 410$\text{--}$14500~$\mathrm{MHz}$ and FR2 NTN from $10.7\text{--}30$~$\mathrm{GHz}$. Several NTN bands overlap in frequency because spectrum allocations differ across regions. This overlap can simplify device design and support multi-region operation. Lower frequencies, such as the L and S bands, have lower propagation loss and are better suited to compact terminals, but they typically offer limited bandwidth. This constrains throughput for data-intensive UAS services, such as real-time video streaming. Although this limited bandwidth may satisfy current ICNS requirements, it may become insufficient as aircraft density increases in the future. Higher-frequency options, such as the Ku and Ka bands, offer larger bandwidths and higher throughput. However, these bands are more sensitive to atmospheric conditions. For example, Starlink operates in the Ku band, where rain and high humidity can reduce performance~\cite{StarlinkWeather}. As a result, operating in these bands typically requires high-gain directional antennas, which can be too large and heavy for small unmanned aircraft. In addition, satellite receivers operating in these bands require high antenna gain to overcome the substantial path loss \cite{ullah3gpp}. Therefore, we consider the L- and S-bands more suitable candidates for D2D links supporting ICNS services.

\begin{table}
   \caption{5G NTN bands for uplink and downlink.}
\label{tab:tab3}
  \begin{tabular}{|m{10.81mm}|b{8.13mm} m{2.15mm} p{14.55mm}|p{8.13mm} m{2.15mm} p{12.23mm}|}
    \hline
    \textbf{NR band} &
    \multicolumn{3}{c|}{\textbf{Uplink}} &
    \multicolumn{3}{c|}{\textbf{Downlink}} \\
    \hline
    % n2523  & $~~~~~200$ & $-$ & $2020$~$\mathrm{MHz}$   & $~~~2180$  & $-$ & $2200$~$\mathrm{MHz}$ \\
    n253 & $~~~1668$  & $-$ & $1675$~$\mathrm{MHz}$   & $~~~1518$  & $-$ & $1525$~$\mathrm{MHz}$ \\
    n254 & $~~~1610$  & $-$ & $1626.5$~$\mathrm{MHz}$ & $2483.5$   & $-$ & $2500$~$\mathrm{MHz}$ \\
    n255 & $1626.5$   & $-$ & $1660.5$~$\mathrm{MHz}$ & $~~~1525$  & $-$ & $1559$~$\mathrm{MHz}$ \\
    n256 & $~~~1980$  & $-$ & $2010$~$\mathrm{MHz}$   & $~~~2170$  & $-$ & $2200$~$\mathrm{MHz}$ \\
    n510 & $~~~~27.5$ & $-$ & $28.35$~$\mathrm{GHz}$  & $~~~~17.7$ & $-$ & $20.2$~$\mathrm{GHz}$ \\
    n511 & $~~28.35$  & $-$ & $30$~$\mathrm{GHz}$     & $~~~~17.7$ & $-$ & $20.2$~$\mathrm{GHz}$ \\
    n512 & $~~~~27.5$ & $-$ & $30$~$\mathrm{GHz}$     & $~~~~17.7$ & $-$ & $20.2$~$\mathrm{GHz}$ \\
    \hline
  \end{tabular}
\end{table}
\section{System Model}
\label{sec:sec3}
\begin{figure}[t!]
\centerline{\includegraphics[width=\columnwidth]{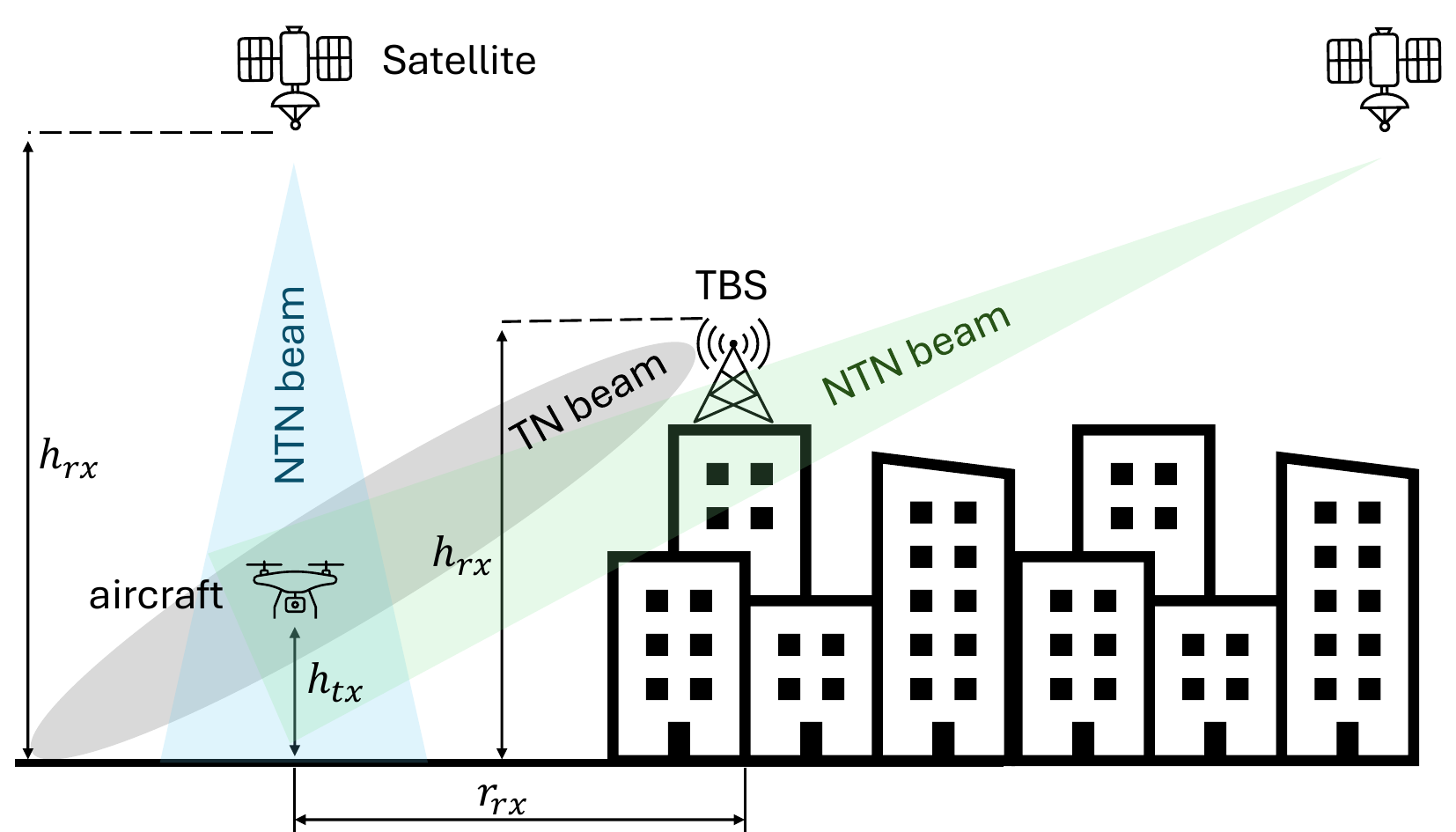}}
\caption{Illustrations of unmanned aircraft operation in urban environment using TN and NTN systems.}
\label{fig:fig1}
\end{figure}

This section presents the mathematical modeling for LoS probability, path loss, vertical radiation pattern of the antennas, and analytical formulation for computing the received signal strength. Fig.~\ref{fig:fig1} presents a high-level illustration of our scenario. Table~\ref{tab:tab4} shows the key notations and their definitions. We investigate an urban scenario in which a TBS is installed on a building rooftop at a height ($h_{rx}$) above ground level~(AGL). An aircraft is located at a height ($h_{tx}$) AGL, at a horizontal distance ($r_{rx}$) from the TBS. We also consider an NTN system with a LEO satellite orbiting the Earth at a height ($h_{rx}$) AGL, whose elevation angle varies between $10^\circ$ and $90^\circ$ due to the satellite's orbital motion.
\begin{table}[t!]
    \centering
    \caption{The list of key notations and their definitions.}
    \resizebox{\columnwidth}{!}{\begin{tabular}{|l|l|}
        \hline
        \textbf{Acronyms} & \textbf{Definition} \\ \hline
$\alpha$ &  Ratio of built-up area to the total site area\\ 
$\beta$ &  Mean building density measured as buildings/$\mathrm{km}$$^2$ \\
$\gamma$ & Building height distribution\\ 
$b_1$& Number of buildings passed per kilometer\\
$b_r$& Number of buildings crossed in ray path\\
%$CL$& Clutter loss\\
%$CL_{max}$ & Maximum clutter loss\\
$d_i$& Buildings distances\\
$h_i$& Ray height at the distance $d_i$\\
$\mathrm{P_L}$  & Total path loss\\
$P_\mathrm{L_{LoS}}$ & LoS path loss\\
$P_\mathrm{L_{NLoS}}$ & NLoS path loss\\
$\delta_r$& Separation between the buildings\\
$\mathbb{P}_{\mathrm{LoS}}$ &Line-of-sight (LoS) probability\\ 
        \hline
    \end{tabular}}
\label{tab:tab4}
\end{table}

\subsection{Line-of-Sight Probability Model}
In this paper, we calculate LoS probability using the ITU-R P.1410-6 statistical model (refer to Section 2.1.4 statistical model in \cite{ITU_R_P1410_2023}). This model requires three parameters, including: ($i$) $\alpha$ defines the ratio of built-up area to the total site area (dimensionless); ($ii$) $\beta$ is the mean building density, measured in buildings per square kilometer, and ($iii$) $\gamma$ defines the building-height distribution, assumed to follow a Rayleigh distribution. If we assume that buildings are placed evenly in a grid, a one-kilometer ray would pass over $\sqrt{\beta}$  buildings. However, since only a ratio $\alpha$ of the land is actually covered with buildings, the number of buildings crossed per kilometer is reduced accordingly as
\begin{equation}
b_1 = \sqrt{\alpha \beta}
\label{eq:1}
\end{equation}
Ray path ($r_{rx}$), also known as ground range, is a two-dimensional horizontal distance between the transmitter and the receiver. The unit of $r_{rx}$ is a kilometer. The total number of buildings crossed over a ray path ($r_{rx}$) is given by
\begin{equation}
b_r = \left\lfloor r_{rx} \, b_1 \right\rfloor
\label{eq:2}
\end{equation}
where $\eqref{eq:2}$ uses the floor function  ($\left\lfloor \right\rfloor$) to ensure that the output is an integer number. In the next step, ITU-R P.1410-6 statistical model \cite{ITU_R_P1410_2023} assumes that the buildings are evenly spaced between the transmitter and the receiver, where each building is located at a distance ($d_i$) as follows
\begin{equation}
d_i = \left(i + \tfrac{1}{2}\right)\,\delta_r, 
\qquad i \in \{0,1,\ldots,(b_r - 1)\}
\label{eq:3}
\end{equation}

where 
$\delta_r = \frac{r_{rx}}{b_r}$ is the separation between the buildings. The variable $d_i$ denotes the horizontal distance from the transmitter to the potential blockage point on the $i$-th building. At this distance, $h_{i}$ corresponds to the height of the ray at the obstruction point as
\begin{equation}
h_i = h_{tx} - \frac{d_i\left(h_{tx} - h_{rx}\right)}{r_{rx}}
\label{eq:4}
\end{equation}

where parameters $h_{tx}$ and $h_{rx}$ denote the heights of the transmitter (aircraft) and receiver (TBS or satellite) AGL, respectively. At a horizontal distance $d_i$, the term $P_i$ represents the probability that the building height does not exceed ray height $h_{i}$, i.e., the probability that the building height is less than ray height $h_{i}$. Thus, the probability $P_i$ is given by

\begin{equation}
P_i = 1 - \exp\left(-\frac{h_i^2}{2\gamma^2}\right)
\label{eq:5}
\end{equation}

The $P_{\mathrm{LoS},i}$ represents probability of having a LoS path at distance $d_i$ as
\begin{equation}
\mathbb{P}_{\mathrm{LoS},i} = \prod_{j=0}^{i} P_j, 
\quad j \in \{0, \ldots, i\}
\label{eq:6}
\end{equation}
\subsection{Path loss Model}
The total path loss is expressed as the weighted combination of LoS and non-line-of-sight (NLoS) components, where the weights correspond to their respective occurrence probabilities as in \cite{Akram}. Specifically, it is given by the sum of the product of the LoS probability and the LoS path loss, as well as the product of the NLoS probability and the NLoS path loss. The total path loss is modeled as
\begin{equation}
\mathrm{P_L} = \mathbb{P}_{\mathrm{LoS}}  \times {P_\mathrm{L_{LoS}}} + (1-\mathbb{P}_{\mathrm{LoS}}) \times P_{\mathrm{L_{NLoS}}}
\label{eq:7}
\end{equation}
Similar to 3GPP TR 38.811 \cite{3gppTR38811}, LoS path loss is defined by
\begin{equation}
{P_\mathrm{L_{LoS}}} = 32.45 + 20\log_{10}(\mathrm{f_{c}}) + 20\log_{10}(\mathrm{r}) + \mathrm{SF} 
\label{eq:8}
\end{equation}
where carrier frequency ($\mathrm{f_{c}}$) in $\mathrm{GHz}$, $\mathrm{r}$ in $\mathrm{m}$ is link distance, also called slant range, and $\mathrm{SF}$ denotes the shadow fading component.  Following 3GPP TR 38.811 \cite{3gppTR38811}, path loss for NLoS ray is calculated as
\begin{equation}
{P_\mathrm{L_{NLoS}}} = 32.45 + 20\log_{10}(\mathrm{f_{c}}) + 20\log_{10}(\mathrm{r}) + \mathrm{SF} + \mathrm{CL}
\label{eq:9}
\end{equation}
where $\mathrm{CL}$ represents the clutter loss. Unlike the path loss for a ray under perfect LoS conditions, NLoS propagation incurs additional attenuation due to clutter loss $\mathrm{CL}$, caused by blockages such as surrounding buildings and ground obstacles. Note that the clutter-loss model in 3GPP TR~38.811~\cite{3gppTR38811} assumes a user at ground level. In this paper, we adopt an altitude-dependent clutter attenuation model to account for the aircraft's variable height, which ranges from 1 to 300~$\mathrm{m}$. 
Particularly, the clutter loss is modeled as an exponentially decaying function of the aircraft height as%\cite{Decay} of the aircraft height as
\begin{equation}
\mathrm{CL} =\mathrm{CL_{max}}\exp(-h_{tx}/h_{0})
\label{eq:10}
\end{equation}

where 
$\mathrm{CL_{max}}$ is the maximum clutter loss, and \mbox{$h_{0} = 1.25\times \gamma$} $\mathrm{m}$ represents building height. In $\eqref{eq:10}$ formulation reflects the reduction of clutter effects as the aircraft rises above the urban infrastructure.

In \eqref{eq:8} and \eqref{eq:9}, $\mathrm{r}$ is link distance (slant range) between transmitter (aircraft) and receiver (TBS or satellite). In case of TN,  link distance between aircraft and TBS is calculated as $\mathrm{r} = \sqrt{ (r_{\mathrm{rx}}\times 1000)^2 + \left| h_{\mathrm{tx}} - h_{\mathrm{rx}} \right|^2 }$. In case of NTN, we consider a satellite as a receiver and aircraft as a transmitter. We calculate link distance using the MATLAB built-in function $\mathrm{r}$ = \verb|height2range(tgtht,anht,el)|\footnote{https://se.mathworks.com/help/radar/ref/height2range.html} which converts a target height to propagated range, where \verb|tgtht| is receiver height as $h_{\mathrm{rx}}$ = 300,000 $\mathrm{m}$, \verb|anht| is variable transmitter (aircraft) height between 1 to 300 $\mathrm{m}$, and \verb|el| is the elevation angle between aircraft and satellite. Similarly, in the case of NTN,  we calculate $r_{\mathrm{rx}}$, which is a two-dimensional distance also called ground range between aircraft and satellite using the MATLAB built-in function as $r_{\mathrm{rx}}$~=~\verb|height2grndrange(tgtht,anht,el)|\footnote{https://se.mathworks.com/help/radar/ref/height2grndrange.html}.

\subsection{Vertical antenna pattern }
%The vertical radiation pattern of an antenna depends on the antenna tilt,  half-power beamwidth, and the minimum vertical gain.
In 3GPP TR 36.814 \cite{3gppTR36814}, the vertical antenna pattern in $\mathrm{dB}$ is defined by the antenna downtilt angle,  half-power beamwidth, and the minimum vertical gain as
\begin{equation}
\mathrm{A_{V(\theta)}} = -\min \left[
12 \left( \frac{\theta - \theta_{\text{etilt}}}{\theta_{3\text{dB}}} \right)^2,
\, SLA_V
\right]
\end{equation}

where $\theta_{3\text{dB}}$ is the half-power beamwidth,  $SLA_V$ in $\mathrm{dB}$ is minimum
vertical gain, $\theta_{\text{etilt}}$ is electrical antenna downtilt angle, and $\theta \in [-90^\circ,\,90^\circ]$ is the elevation angle. However, the actual angle experienced by the aircraft depends on TBS height $h_{\mathrm{rx}}$, aircraft height $h_{\mathrm{tx}}$, and link distance $\mathrm{r}$. We calculate angle $\theta$ using the MATLAB built-in function \verb|height2el|\footnote{https://se.mathworks.com/help/radar/ref/height2el.html}.

\subsection{Received signal power}
We use the Friis transmission formula \cite{Friis} to calculate the received power in $\mathrm{dB}$ as
\begin{equation}
\mathrm{P_{rx}} = \mathrm{P_{tx}} + \mathrm{G_{tx}} + \mathrm{G_{rx}} + \mathrm{A_{V}} -\mathrm{P_{L}}
\end{equation}

where $\mathrm{P_{tx}}$, $\mathrm{G_{tx}}$, $\mathrm{G_{rx}}$, $\mathrm{A_{V}}$, and $\mathrm{P_{L}}$ denote the transmit power, transmitter antenna gain, receiver antenna gain, vertical antenna gain, and path loss, respectively. Horizontal antenna pattern is not explicitly modeled; instead, we consider a best-case condition assuming TBS to be horizontally omnidirectional as in \cite{Cherif}. For the NTN scenario, the satellite antenna radiation pattern is assumed to remain uniform within a given spot beam and independent of the aircraft altitude. Consequently, the vertical antenna pattern $A_{V}$ is neglected in the NTN analysis. 

\section{Results}
\label{sec:sec4}
%\subsection{Key parameters and assumptions}
\subsection{Evaluation setup}
Table~\ref{table:tab5} lists the key parameters and their values used in our evaluation. Specifically, we consider an urban scenario with $\alpha=0.3$, $\beta=500$, and $\gamma=15$~\cite{Akram}. We consider a 6G UE with D2D capabilities mounted on the aircraft, transmitting at 23~$\mathrm{dBm}$ and using an antenna with 0~$\mathrm{dBi}$ gain. We assume the TBS is installed at the height ($h_{rx}$) of 25~$\mathrm{m}$. TBS carrier frequency ($\mathrm{f_{c}}$) is 3.6~$\mathrm{GHz}$ and it features a receiving antenna with gain $\mathrm{G_{rx}}$~=~8~$\mathrm{dBi}$, half-power beamwidth ($\theta_{3\text{dB}}$)~=~$10^\circ$, electrical antenna downtilt ($\theta_{\text{etilt}}$)~=~$6^\circ$, and minimum vertical gain ($SLA_V$)~=~20~$\mathrm{dB}$~\cite{3gppTR36814,Cherif}. We consider the reference receiver sensitivity is approximately -100 $\mathrm{dBm}$ for operation at $\mathrm{f_{c}}$ = 3.6 $\mathrm{GHz}$ with 5 $\mathrm{MHz}$ channel bandwidth and 15 $\mathrm{kHz}$ subcarrier spacing (SCS)~\cite{3gppTS38521_1}. 

 For the NTN segment, we assume a satellite orbiting at height ($h_{rx}$) of 300,000 $\mathrm{m}$, which features a higher antenna gain $\mathrm{G_{rx}}$~=~38~$\mathrm{dBi}$ to compensate for extreme path loss. Recent advancements in satellite communications have demonstrated the feasibility of high-gain satellite antennas for D2D connectivity~\cite{ullah3gpp}. For example, AST SpaceMobile is developing LEO satellite systems capable of directly connecting to standard 5G mobile devices without hardware modification. These systems operate in frequency bands comparable to conventional terrestrial cellular bands. Reported uplink frequencies include 846.5–849 $\mathrm{MHz}$, 845–846.5 $\mathrm{MHz}$, and 788–798 $\mathrm{MHz}$, while downlink operation includes 891.5–894 $\mathrm{MHz}$, 890–891.5 $\mathrm{MHz}$, and 758–768 $\mathrm{MHz}$. The satellite payload employs electronically steerable beams and achieves a maximum antenna gain on the order of 36 $\mathrm{dBi}$~\cite{FCC_AST_Application}.

Similarly, Starlink has initiated direct-to-cell services supporting 4G connectivity in collaboration with T-Mobile, utilizing spectrum in the 1910–1915 $\mathrm{MHz}$ and 1990–1995 $\mathrm{MHz}$ bands. Publicly available information indicates peak satellite antenna gains of approximately 38 $\mathrm{dBi}$ for such systems~\cite{SpaceX_TMobile_Technical_Narrative}. In addition, system-level evaluations reported in 3GPP TR 38.821~\cite{3gppTR38821} consider satellite antenna gains as high as 51 $\mathrm{dBi}$ for S-band (2 $\mathrm{GHz}$) NTN scenarios. These examples support the assumption of high satellite antenna gain in our link budget analysis. Furthermore, in the NTN case, we set the receiver sensitivity threshold for a channel with 5 $\mathrm{MHz}$ bandwidth at 15 kHz subcarrier spacing (SCS), which is -102.4~$\mathrm{dBm}$ as in 3GPP~TS~38.108~\cite{3gppTS38108}. 

\begin{table}[!t]
\caption{Key parameters used in the evaluation.}
\centering
\small
\begin{tabular}{@{}ll@{}}
\hline
\textbf{Parameter} & \textbf{Value} \\
\hline

\multicolumn{2}{c}{\textbf{Common Parameters}} \\
\hline
$\alpha$ & 0.3 \\
$\beta$ & 500 buildings/$\mathrm{km}$$^2$ \\
$\gamma$ & 15 $\mathrm{m}$ \\
Aircraft height ($h_{tx}$) & 1--300 $\mathrm{m}$ \\
Bandwidth & 5 $\mathrm{MHz}$ \\
LoS shadow fading ($\mathrm{SF}$) & 4 $\mathrm{dB}$\\
Maximum clutter loss ($CL_{\max}$) & 34.3 $\mathrm{dB}$ \\
NLoS shadow fading ($\mathrm{SF}$) & 6 $\mathrm{dB}$\\
Receiver sensitivity & -$102.4$ $\mathrm{dBm}$ \\
UE antenna gain ($G_{tx}$) & 0 $\mathrm{dBi}$ \\
UE transmit power ($P_{tx}$) & 23 $\mathrm{dBm}$ \\
\hline

\multicolumn{2}{c}{\textbf{TN Parameters}} \\
\hline
Carrier frequency ($\mathrm{f_{c}}$) & 3.6 $\mathrm{GHz}$ \\
Electrical antenna downtilt ($\theta_{\text{etilt}}$) & $6^\circ$ \\
TBS antenna gain ($G_{rx}$) & 8 $\mathrm{dBi}$ \\
TBS height ($h_{rx}$) & 25 $\mathrm{m}$ \\
TBS half-power beamwidth ($\theta_{3\text{dB}}$) & $10^\circ$ \\
%\textcolor{red}{Link distance} ($\mathrm{d}$) & \\
Vertical antenna gain ($SLA_V$) & 20 $\mathrm{dB}$ \\
\hline
\multicolumn{2}{c}{\textbf{NTN Parameters}} \\
\hline
Carrier frequency ($\mathrm{f_{c}}$) & 2 $\mathrm{GHz}$ \\
Receiver antenna gain ($G_{rx}$) & 38 $\mathrm{dBi}$\\
Satellite altitude ($h_{rx}$)& 300 $\mathrm{km}$ \\
Satellite elevation angles & $10^\circ$, $30^\circ$, $90^\circ$ \\
%Satellite link distance ($\mathrm{d}$) & 1237,  572, 300 km\\
Satellite orbit & Very LEO \\

\hline

\end{tabular}
\label{table:tab5}
\end{table}

\subsection{Numerical results}
\begin{figure}[t!]
\centerline{\includegraphics[width=\columnwidth]{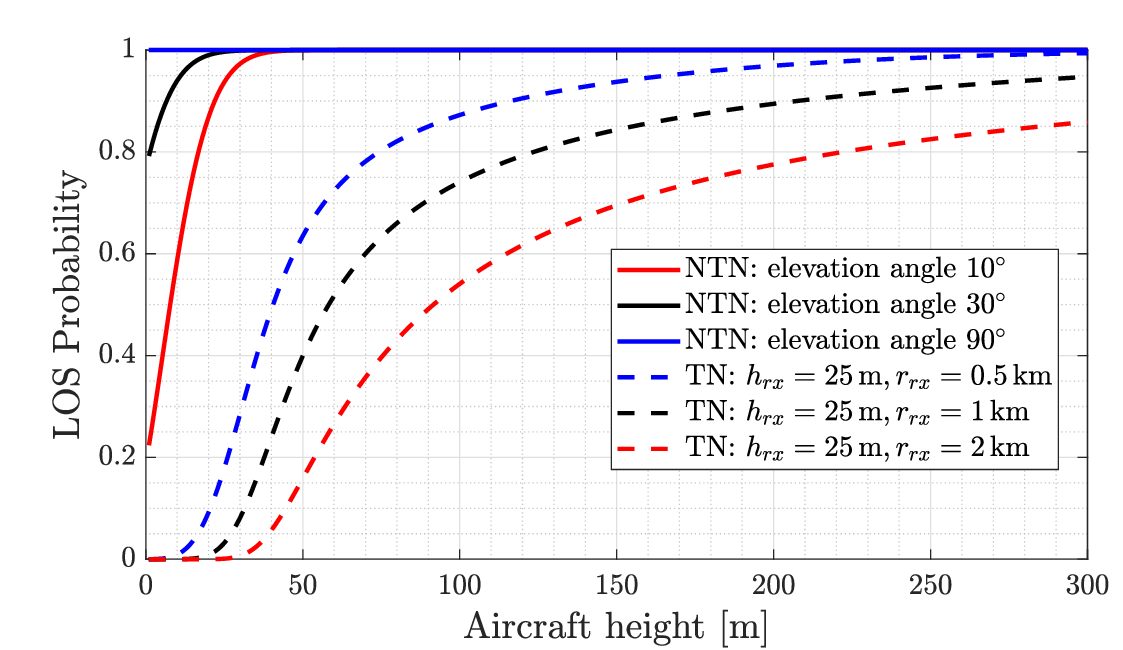}}
\caption{LoS probability as a function of aircraft height above the ground.}
\label{fig:fig2}
\end{figure}

In our evaluation, we consider a range of scenarios for the TN and NTN segments (see Fig.~\ref{fig:fig1}). For the TN case, we evaluate cases with several horizontal distances \mbox{$r_{\mathrm{rx}} \in$ [0.5, 1, 2]~$\mathrm{km}$}, while for the NTN case, we consider elevation angles \verb|el| $\in$ [$10^\circ, 30^\circ, 90^\circ$] between the aircraft and the satellite, covering a wide range of possible configurations.

\textit{LoS probability:} Fig.~\ref{fig:fig2} illustrates the LoS probability for TN and NTN connections as a function of the aircraft height ($h_{tx}$). The TN link exhibits a significantly lower LoS probability compared with all three NTN cases corresponding to elevation angles of 10$^\circ$, 30$^\circ$, and 90$^\circ$. This difference arises because, in the TN scenario, the TBS is located at a height of $h_{rx}$ = 25 $\mathrm{m}$ AGL, whereas the NTN scenario considers a LEO satellite at an altitude of $h_{rx}$ = 300{,}000$~\mathrm{m}$. Consequently, the link distance (slant range), denoted by $\mathrm{r}$, between the aircraft and the satellite encounters fewer obstacles than in the case of communication with a TBS. As the aircraft moves farther from the TBS, the horizontal distance $r_{rx}$ increases, which reduces the ray slope and increases the number of blockages, thereby reducing the LoS probability. Specifically, in the TN case, when $r_{rx}$ = 0.5 $\mathrm{km}$, the LoS probability is approximately 0.5 at an aircraft height of 40 $\mathrm{m}$. In contrast, for $r_{rx}$ = 2 $\mathrm{km}$, the LoS probability remains close to zero until the aircraft height ($h_{tx}$) reaches about 30 $\mathrm{m}$, and it increases to 0.5 at \mbox{$h_{tx}$ = 90 $\mathrm{m}$}. At the maximum aircraft height of 300 $\mathrm{m}$, the LoS probabilities are approximately 0.99, 0.94, and 0.85 for $r_{rx}$ = 0.5, 1, and 2 $\mathrm{km}$, respectively.

In the NTN scenario, when the aircraft height is 10 $\mathrm{m}$, the LoS probabilities are approximately 0.58, 0.94, and 1 for elevation angles of 10$^\circ$, 30$^\circ$, and 90$^\circ$, respectively. The LoS probability is close to one when the aircraft height exceeds 40~$\mathrm{m}$ AGL for all three NTN cases.
\begin{figure}[t!]
\centerline{\includegraphics[width=\columnwidth]{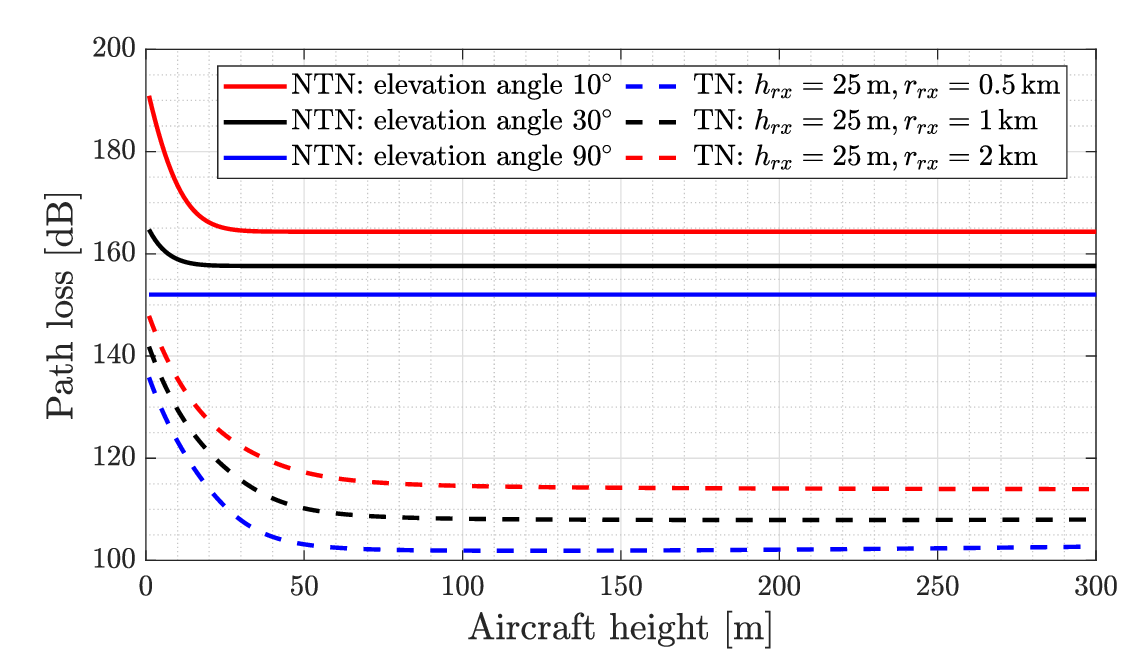}}
\caption{Path loss as a function of aircraft height above the ground level.}
\label{fig:fig3}
\end{figure}

\textit{Path loss:} We further use the LoS results from Fig.~\ref{fig:fig2} to derive the path loss as a function of the aircraft height AGL, using \eqref{eq:7}. Fig.~\ref{fig:fig3} illustrates the resulting path loss. As expected, the TN connection exhibits significantly lower path loss than the NTN connection due to the shorter link distance $\mathrm{r}$ between the aircraft and the TBS. Specifically, when the aircraft is close to ground level, the TN link experiences path losses of 135~$\mathrm{dB}$, 141~$\mathrm{dB}$, and 147~$\mathrm{dB}$ for distances between the aircraft and TBS of $r_{rx}$ = 0.5, 1, and 2 $\mathrm{km}$, respectively. The path loss decreases sharply to 103~$\mathrm{dB}$, 110~$\mathrm{dB}$, and 117~$\mathrm{dB}$ as the aircraft height increases to 50~$\mathrm{m}$.

In the NTN case, a LEO satellite at an altitude of \mbox{$h_{rx}$ = 300{,}000 $\mathrm{m}$} is considered, resulting in link distances $\mathrm{r}$ of 1237 $\mathrm{km}$, 572 $\mathrm{km}$, and 300 $\mathrm{km}$ at elevation angles of 10$^\circ$, 30$^\circ$, and 90$^\circ$, respectively. At an elevation of 90$^\circ$, the satellite is assumed to be directly above the aircraft, resulting in a perfect LoS condition; hence, according to our model the path loss remains constant at 152~$\mathrm{dB}$ for all aircraft heights. In contrast, the NTN case with an elevation angle of 10$^\circ$ exhibits a path loss exceeding 174~$\mathrm{dB}$ when the aircraft height is 10~$\mathrm{m}$, where the NLoS component is dominant, leading to high clutter losses. After the aircraft height exceeds 30~$\mathrm{m}$, the path loss stabilizes at approximately 164~$\mathrm{dB}$. For the NTN scenario with an elevation angle of 30$^\circ$, the path loss is 158~$\mathrm{dB}$ when the aircraft altitude reaches 10~$\mathrm{m}$. To compensate for these high path losses, we assume that the satellite has a higher receive antenna gain of $\mathrm{G_{rx}}$~=~38~$\mathrm{dBi}$.

\begin{figure}[t!]
\centerline{\includegraphics[width=\columnwidth]{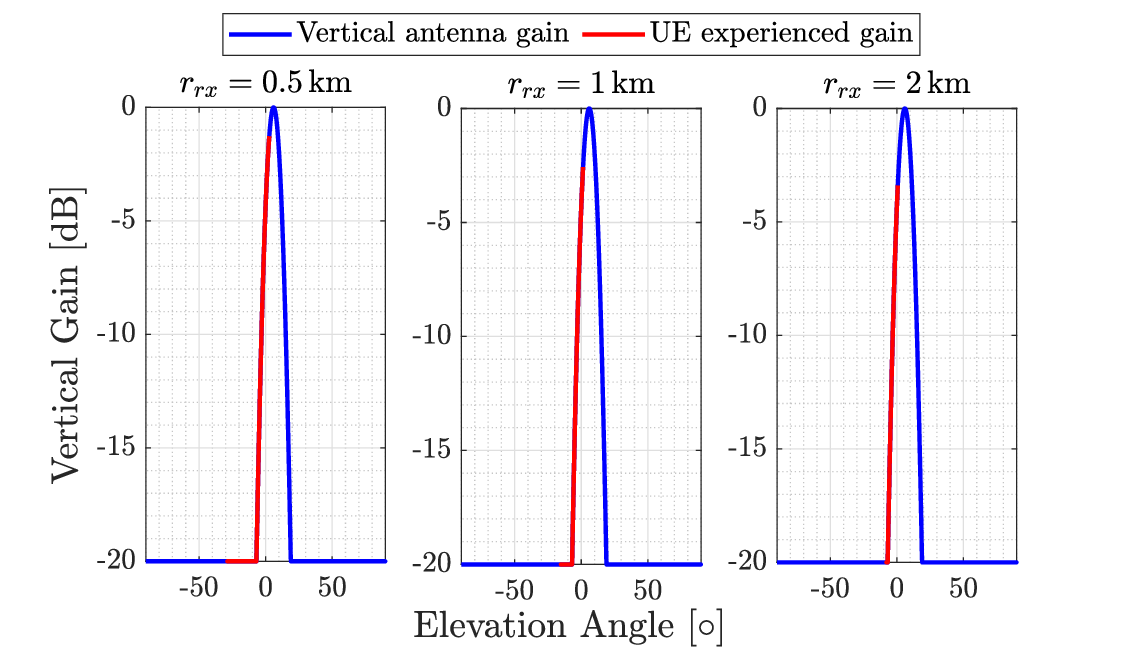}}
\caption{Impact of down tilting on the vertical antenna gain.}
\label{fig:fig4}
\end{figure}

\textit{Vertical antenna gain:} As mentioned earlier, the coverage of the cellular TBS is optimized to serve users at ground level~\cite{3gppTR36814}. This is achieved through antenna down-tilting~\cite{Cherif}. Fig. \ref{fig:fig4} shows the vertical antenna gain pattern when the electrical antenna downtilt $\theta_{\text{etilt}}$ = $6^\circ$. In Fig. \ref{fig:fig4}, the blue curve represents the overall vertical antenna gain pattern, while the red curve shows the gain experienced by a UE mounted on the aircraft. It can be observed that the magnitude of the experienced gain decreases as the aircraft moves farther away from the TBS.

\begin{figure}[t!]
\centerline{\includegraphics[width=\columnwidth]{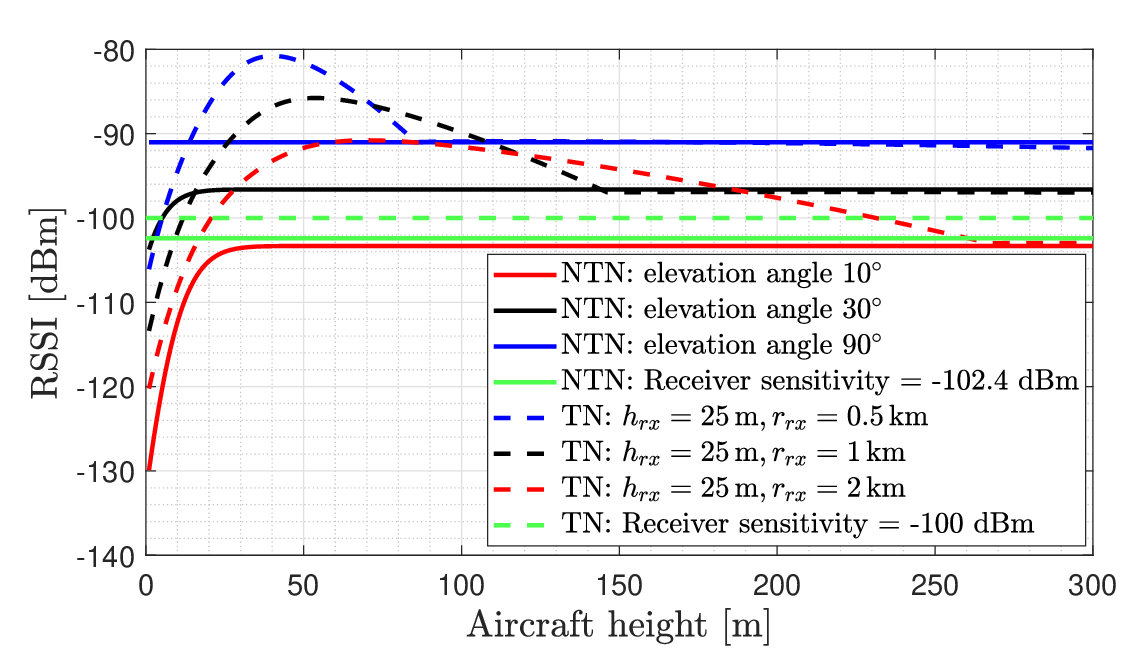}}
\caption{Received signal strength indicator as a function of aircraft height.}
\label{fig:fig5}
\end{figure}
\textit{Coverage:} Note that, for a channel bandwidth of 5 $\mathrm{MHz}$ and SCS of 15 kHz, the receiver sensitivity requirements are -100 $\mathrm{dBm}$ and -102.4 $\mathrm{dBm}$ for TN and NTN connections, respectively. A RSSI above these receiver sensitivity thresholds is required to successfully establish a connection between the UE and the TBS or satellite. Fig. \ref{fig:fig5} illustrates the RSSI as a function of aircraft height for both TN and NTN connections. When the aircraft is at $r_{rx}$ = 0.5, 1, and 2~$\mathrm{km}$ from the TBS, a TN connection is feasible only when the aircraft altitude $h_{tx}$ exceeds 5, 12, and 20~$\mathrm{m}$, respectively. For $r_{rx} = 2~\mathrm{km}$, the link between the aircraft and the TBS enters outage when $h_{tx}$ goes above 230~$\mathrm{m}$.

In the NTN case, connectivity at an elevation angle of 90$^\circ$ is feasible regardless of aircraft height, while at 30$^\circ$ it becomes feasible when the aircraft height exceeds 3 $\mathrm{m}$. However, the NTN link with an elevation angle of 10$^\circ$ fails to establish a connection across the entire aircraft height range. Communication at 10$^\circ$ can become feasible above 25 $\mathrm{m}$ if the satellite receiving antenna gain $\mathrm{G_{rx}}$ is further increased by 2 $\mathrm{dBi}$ to 40 $\mathrm{dBi}$.

\section{Discussion}
\label{sec:sec5}
 As described in Section~\ref{sec:sec2b}, TNs typically serve as the primary connectivity solution in areas with ground infrastructure. On the other hand, NTNs, especially satellites, extend coverage to remote and oceanic areas. Beyond coverage extension, NTNs can also complement TNs in dense urban environments by managing excess traffic during peak-demand periods. Furthermore, in the event of terrestrial network failures, satellite links can maintain service continuity, thereby improving overall system resilience.

As illustrated in Fig.~\ref{fig:fig5}, for a TN with a network density that results in the horizontal distance from any TBS to aircraft less than or equal to $r_{rx}$ = 0.5 $\mathrm{km}$, the TN coverage would ensure an RSSI that is above the minimum receiver sensitivity threshold. Since most urban commercial TNs will have high TBS densities, which would result in much shorter horizontal distances between aircraft and TN base stations, the general use of TNs for low altitude below 300 $\mathrm{m}$ would become feasible for primary connectivity purposes. The access to multiple TBSs can be provisioned via access roaming agreements with multiple local or regional TN service providers, such as commercial Mobile Network Operators (MNOs). Through agreements with multiple NTNs and multiple MNOs, a very resilient combined NTN and TN system can be feasibly established.

Beyond densified TN coverage areas, where TBS may be more than $r_{rx}$ = 0.5 $\mathrm{km}$ of horizontal distance from an aircraft, Fig.~\ref{fig:fig5} also shows the potential for NTN to provide the RSSI required to meet or exceed the minimum receiver sensitivity threshold, when the elevation angle of an aircraft to a serving satellite is near 30$^\circ$ or higher. This shows the feasibility of a combined TN and NTN system to both provide extended coverage where TN coverage is limited or unavailable, and it shows that a combined TN and NTN system can support an increased volume of aircraft via the enhancement (augmentation) of TN capacity with NTN capacity. State-of-the-art commercial mobile networks often have network density that ensures distances to macro or micro base stations are within hundreds of meters. Therefore, the overall results demonstrate that the combined TN and NTN system can provide the extended coverage required to deliver ICNS services at low altitudes. 
\section{Conclusion}
\label{sec:sec6}
This paper presents a vision for integrated Communication, Navigation, and Surveillance relying on hybrid TN–NTN connectivity in 6G networks. This concept assumes that a 6G user equipment with satellite-based direct-to-device capabilities is mounted on an aircraft, enabling it to leverage both terrestrial and non-terrestrial network resources. This can add significant system redundancy, ensuring overall network availability and system resilience. The combined use of terrestrial and non-terrestrial networks enhances coverage availability for CNS services at altitudes where TN coverage is limited or unavailable.  This paper concludes that a UE with D2D capability and the ability to smoothly roam between TN and NTN can provide a feasible solution for using combined networks to support CNS services at low altitude. The results of the modeling and simulation illustrate that a combined TN and NTN system, particularly one operating in an urban environment, can enable low-altitude CNS services to be provisioned with the primary connectivity, extended coverage, and resilience needed to deliver connectivity for ICNS services for low-altitude operation. In future work, we plan to investigate satellite-based device-to-device communication to enable aircraft-to-aircraft and aircraft-to-everything communication via satellite, without relying on feeder links or ground stations. As another future research direction, we plan to investigate the feasibility of extending terrestrial base-station coverage to low-altitude aerial corridors using up-tilted antennas. These aerial corridors are defined by aviation authorities. Our goal is to optimize network performance specifically within these predefined corridors, rather than extending coverage everywhere. This targeted approach will focus on improving the quality and reliability of ICNS services for aerial operations along designated aerial corridors. Our NTN results indicate that communication with LEO satellites is not feasible at low elevation angles. In the future, we also aim to conduct a comprehensive study of this issue by analyzing different satellite constellation densities, ranging from a few hundred to several thousand satellites. This investigation will help us better understand the impact of constellation size on link feasibility and overall system performance. %\textcolor{red}{Asad: Please list a few more ideas for future work}
\if{0}
As described in section~\ref{sec:sec2b}, TNs typically serve as the primary connectivity solution in areas where ground infrastructure is available. On the other hand, NTNs, especially satellites, extend coverage to remote and oceanic areas. Beyond coverage extension, NTNs can also complement TNs in dense urban environments by managing excess traffic during peak-demand periods. Furthermore, in the event of terrestrial network failures, satellite links can maintain service continuity, thereby improving overall system resilience [1].

As illustrated in Fig.~\ref{fig:fig5}, for a TN with a network density that results in the horizontal distance from any TBS to aircraft less than or equal to $r_{rx} = 0.5$ $\mathrm{km}$, the TN coverage would ensure an RSSI that is above the minimum receiver sensitivity threshold. Since most urban commercial TNs will have high TBS densities, which would result in much shorter horizontal distances between aircraft and TN base stations, the general use of TNs for low altitude below 300 $\mathrm{m}$ would become feasible for primary connectivity purposes. Access to multiple TBS can be provisioned via access roaming agreements with multiple local or regional TN service providers, such as commercial Mobile Network Operators (MNOs). Through agreements with multiple NTNs and multiple MNOs, a very resilient combined NTN and TN system can be feasibly established.

Beyond densified TN coverage areas, where TBS may be more than $r_{rx} = 0.5$ $\mathrm{km}$ of horizontal distance from an aircraft, Fig.~\ref{fig:fig5} also shows the potential for NTN to provide the RSSI required to meet or exceed the minimum receiver sensitivity threshold, when the elevation angle of an aircraft to a serving satellite is near 30$^\circ$ or higher. This shows the feasibility of a combined TN and NTN system to both provide extended coverage where TN coverage is limited or unavailable and this shows the ability of a combined TN and NTN system can support an increased volume of aircraft via the enhancement (augmentation) of TN capacity with NTN capacity. State-of-the-art commercial mobile networks often have network density that ensures distances to macro or micro base stations are within hundreds of meters. Therefore, the overall results demonstrate that the combined TN and NTN system can provide the extended coverage required to deliver ICNS services at low altitudes. %\textcolor{red}{While the results shown in Fig.~\ref{fig:fig5} show that, due to ground clutter loss considered within the path loss modeling described in section III. SYSTEM MODEL would mitigate the ability for NTNs to meet or exceed the minimum receiver sensitivity threshold (-102.4 dBm) where the horizontal distance from base stations is less than 0.5 kilometers, and when the elevation angle of a signal beam to a serving satellite is less than (<) 10 degrees, the TN would be the primary connectivity in such dense urban areas. }

%Regarding system resilience, TABLE 2 in section II. TECHNICAL BACKGROUND AND RELATED WORKS of this paper list multiple NTNs (satellite constellations) and CNS services they provide. And, Section II. B. discusses future communication systems’ D2D capabilities, which will allow standard user equipment to transition seamlessly between terrestrial and satellite links without requiring dedicated satellite hardware [15], [16]. This capability significantly simplifies the use of combined TN and NTN systems for aircraft connectivity, especially UAS, and broadens the applicability of hybrid architectures for low-altitude CNS services. The multiple NTNs and the D2D capabilities enable aircraft to connect to multiple NTNs. This can add significant system redundancy, ensuring overall network availability and system resilience. Similarly, access to multiple TNs can be provisioned via access roaming agreements with multiple local or regional TN service providers, such as commercial Mobile Network Operators (MNOs). Through agreements with multiple NTNs and multiple MNOs, a very resilient combined NTN and TN system can be feasibly established.
\fi

\section*{Acknowledgment}
This work has been performed under the ANTENNAE project, which has received funding from SESAR3 Joint Undertaking under the European Union's HORIZON-SESAR-2023-DES-ER-02 topic, Grant Agreement No. 101167288. However, the views and opinions expressed are those of the authors only and do not necessarily reflect those of the European Union or SESAR 3 JU. Neither the European Union nor the SESAR 3 JU can be held responsible for them. The authors acknowledge the use of ChatGPT for proofreading and grammatical improvements in the preparation of this manuscript. %We used ChatGPT solely for proofreading and improving grammar.
\balance
\bibliographystyle{IEEEtran}
\bibliography{IEEEabrv,reference}

% Generated by IEEEtran.bst, version: 1.14 (2015/08/26)
\begin{thebibliography}{10}
\providecommand{\url}[1]{#1}
\csname url@samestyle\endcsname
\providecommand{\newblock}{\relax}
\providecommand{\bibinfo}[2]{#2}
\providecommand{\BIBentrySTDinterwordspacing}{\spaceskip=0pt\relax}
\providecommand{\BIBentryALTinterwordstretchfactor}{4}
\providecommand{\BIBentryALTinterwordspacing}{\spaceskip=\fontdimen2\font plus
\BIBentryALTinterwordstretchfactor\fontdimen3\font minus \fontdimen4\font\relax}
\providecommand{\BIBforeignlanguage}[2]{{%
\expandafter\ifx\csname l@#1\endcsname\relax
\typeout{** WARNING: IEEEtran.bst: No hyphenation pattern has been}%
\typeout{** loaded for the language `#1'. Using the pattern for}%
\typeout{** the default language instead.}%
\else
\language=\csname l@#1\endcsname
\fi
#2}}
\providecommand{\BIBdecl}{\relax}
\BIBdecl

\bibitem{Giovanni}
M.~Benzaghta, G.~Geraci, R.~Nikbakht, and D.~López-Pérez, ``{UAV} communications in integrated terrestrial and non-terrestrial networks,'' in \emph{GLOBECOM 2022 - 2022 IEEE Global Communications Conference}, 2022, pp. 3706--3711.

\bibitem{Chen}
L.~Chen, M.~A. Kishk, and M.-S. Alouini, ``Dedicating cellular infrastructure for aerial users: Advantages and potential impact on ground users,'' \emph{IEEE Transactions on Wireless Communications}, vol.~22, no.~4, pp. 2523--2535, 2023.

\bibitem{Leon}
V.~Leon, I.~Christofilos, A.~Nesiadis, I.~Paraskevas, J.~Perrela, G.~Ioannopoulos, A.~Tasoulis-Nonikas, M.~Bernou, and J.~Reading, ``Towards enabling {5G-NTN} satellite communications for manned and unmanned rotary wing aircraft,'' in \emph{2024 IEEE 29th International Workshop on Computer Aided Modeling and Design of Communication Links and Networks (CAMAD)}, 2024, pp. 1--6.

\bibitem{figaro2026}
\BIBentryALTinterwordspacing
M.~Figaro, F.~Rossato, A.~Bonora, M.~Giordani, G.~Schembra, and M.~Zorzi, ``Experimental evaluation of a {UAV}-mounted {LEO} satellite backhaul for emergency connectivity,'' 2026. [Online]. Available: \url{https://arxiv.org/abs/2601.03958}
\BIBentrySTDinterwordspacing

\bibitem{Araniti}
G.~Araniti, A.~Iera, S.~Pizzi, and F.~Rinaldi, ``Toward {6G} non-terrestrial networks,'' \emph{IEEE Network}, vol.~36, no.~1, pp. 113--120, 2022.

\bibitem{pasandi2024survey}
H.~B. Pasandi, J.~A. Fraire, S.~Ratnasamy, and H.~Rivano, ``A survey on direct-to-device satellite communications: Advances, challenges, and prospects,'' in \emph{Proceedings of the 2nd International Workshop on LEO Networking and Communication}, 2024, pp. 7--12.

\bibitem{ullah3gpp}
M.~Asad~Ullah, C.~D. Gedara, A.~Anttonen, A.~Sethi, M.~D. Khattak, M.~H{\"o}yhty{\"a}, J.~Wigard, and K.~Mikhaylov, ``{3GPP RedCap} non-terrestrial networks: Direct-to-device feasibility study,'' \emph{Authorea Preprints}, 2025.

\bibitem{ESA_D2D_2026}
{European Space Agency}, ``Direct-to-device {(D2D)} satellite communications – space for {5G/6G} \& sustainable connectivity,'' \url{https://connectivity.esa.int/artes-4-0-programme-overview/space-for-5g-6g-sustainable-connectivity/direct-to-device-d2d}, 2026, accessed: 2026-03-02.

\bibitem{AsadCNS}
M.~Asad~Ullah \emph{et~al.}, ``{5G} integrated communications, navigation, and surveillance: A vision and future research perspectives,'' in \emph{2025 Integrated Communications, Navigation and Surveillance Conference (ICNS)}, 2025, pp. 1--13.

\bibitem{Alshaer}
H.~Alshaer, A.~Ganau, D.~Brilhante, C.~Cleary, M.~A. Ullah, and V.~Kramar, ``Next-generation integrated communications, navigation, and surveillance services,'' in \emph{2025 Integrated Communications, Navigation and Surveillance Conference (ICNS)}, 2025, pp. 1--10.

\bibitem{3gppTR38914}
\BIBentryALTinterwordspacing
{3rd Generation Partnership Project (3GPP)}, ``Study on {6G} scenarios and requirements,'' 3GPP, Technical Report (TR) 38.914, 2025, draft, Release 19. [Online]. Available: \url{https://portal.3gpp.org/desktopmodules/Specifications/SpecificationDetails.aspx?specificationId=4392}
\BIBentrySTDinterwordspacing

\bibitem{3gppTR22870}
\BIBentryALTinterwordspacing
------, ``Study on {6G} use cases and service requirements,'' 3GPP, Technical Report (TR) 22.870, 2025, draft, Release 20. [Online]. Available: \url{https://portal.3gpp.org/desktopmodules/Specifications/SpecificationDetails.aspx?specificationId=4374}
\BIBentrySTDinterwordspacing

\bibitem{3gppTR38811}
\BIBentryALTinterwordspacing
------, ``{3GPP TR 38.811: Study on New Radio (NR) to support non-terrestrial networks},'' 3GPP, Technical Report 38.811, 2020, release 15. [Online]. Available: \url{https://portal.3gpp.org/desktopmodules/Specifications/SpecificationDetails.aspx?specificationId=3234}
\BIBentrySTDinterwordspacing

\bibitem{3gppTR36814}
------, ``{3GPP TR 36.814: Evolved Universal Terrestrial Radio Access (E-UTRA); Further advancements for E-UTRA physical layer aspects},'' \url{https://portal.3gpp.org/desktopmodules/Specifications/SpecificationDetails.aspx?specificationId=2493}, 3GPP, Technical Report 36.814, 2010, release 9.

\bibitem{3gppTS38108}
\BIBentryALTinterwordspacing
------, ``{NR; Satellite Access Node Radio Transmission and Reception},'' 3GPP, Technical Specification (TS) 38.108, 2022, release 17, under change control. [Online]. Available: \url{https://portal.3gpp.org/desktopmodules/Specifications/SpecificationDetails.aspx?specificationId=3934}
\BIBentrySTDinterwordspacing

\bibitem{3gppTR36777}
\BIBentryALTinterwordspacing
------, ``{3GPP TR 36.777: Study on Enhanced LTE Support for Aerial Vehicles},'' 3GPP, Technical Report 36.777, 2018, release 15. [Online]. Available: \url{https://www.3gpp.org/ftp/Specs/archive/36_series/36.777/}
\BIBentrySTDinterwordspacing

\bibitem{Abdalla}
A.~S. Abdalla and V.~Marojevic, ``Communications standards for unmanned aircraft systems: The {3GPP} perspective and research drivers,'' \emph{{IEEE Communications Standards Magazine}}, vol.~5, no.~1, pp. 70--77, 2021.

\bibitem{khuwaja}
A.~A. Khuwaja, Y.~Chen, N.~Zhao, M.-S. Alouini, and P.~Dobbins, ``{A Survey of Channel Modeling for UAV Communications},'' \emph{{IEEE Communications Surveys \& Tutorials}}, vol.~20, no.~4, pp. 2804--2821, 2018.

\bibitem{Alhourani}
A.~Al-Hourani, ``On the probability of line-of-sight in urban environments,'' \emph{IEEE Wireless Communications Letters}, vol.~9, no.~8, pp. 1178--1181, 2020.

\bibitem{Gapeyenko}
M.~Gapeyenko, D.~Moltchanov, S.~Andreev, and R.~W. Heath, ``Line-of-sight probability for {mmWave}-based uav communications in {3D} urban grid deployments,'' \emph{{IEEE Transactions on Wireless Communications}}, vol.~20, no.~10, pp. 6566--6579, 2021.

\bibitem{Cui}
Z.~Cui \emph{et~al.}, ``Coverage analysis of cellular-connected uav communications with 3gpp antenna and channel models,'' in \emph{2021 IEEE Global Communications Conference (GLOBECOM)}, 2021, pp. 1--6.

\bibitem{Cherif}
N.~Cherif and Q.-U.-A. Nadeem, ``Merits of serving uavs via terrestrial networks: A vertical antenna radiation study,'' in \emph{ICC 2025 - IEEE International Conference on Communications}, 2025, pp. 1470--1475.

\bibitem{Nils}
\BIBentryALTinterwordspacing
N.~Mäurer, T.~Guggemos, T.~Ewert, T.~Gräupl, C.~Schmitt, and S.~Grundner-Culemann, ``Security in digital aeronautical communications a comprehensive gap analysis,'' \emph{International Journal of Critical Infrastructure Protection}, vol.~38, p. 100549, 2022. [Online]. Available: \url{https://www.sciencedirect.com/science/article/pii/S187454822200035X}
\BIBentrySTDinterwordspacing

\bibitem{Arroyo}
A.~Ramírez-Arroyo, T.~B. Sørensen, and P.~Mogensen, ``Terrestrial {5G and Starlink} ntn multi-connectivity toward {6G} communications integration era: An empirical assessment,'' \emph{IEEE Open Journal of the Communications Society}, vol.~6, pp. 5269--5283, 2025.

\bibitem{rinaldi2020non}
F.~Rinaldi, H.-L. Maattanen, J.~Torsner, S.~Pizzi, S.~Andreev, A.~Iera, Y.~Koucheryavy, and G.~Araniti, ``Non-terrestrial networks in {5G} \& beyond: A survey,'' \emph{IEEE access}, vol.~8, pp. 165\,178--165\,200, 2020.

\bibitem{Taha}
H.~Taha, P.~Vári, and A.~Lapsánszky, ``Direct-to-device satellite communications in the {European Union}: Spectrum allocation and regulatory pathways within the {ITU} framework,'' \emph{IEEE Access}, vol.~13, pp. 190\,556--190\,581, 2025.

\bibitem{Feng}
F.~Wang, S.~Zhang, E.-K. Hong, and T.~Q.~S. Quek, ``Constellation as a service: Tailored connectivity management in direct-satellite-to-device networks,'' \emph{IEEE Communications Magazine}, vol.~63, no.~11, pp. 30--36, 2025.

\bibitem{9074973}
R.~Dilli, ``Analysis of {5G} wireless systems in {FR1} and {FR2} frequency bands,'' in \emph{2020 2nd International Conference on Innovative Mechanisms for Industry Applications (ICIMIA)}, 2020, pp. 767--772.

\bibitem{11211403}
S.~J. Seah, S.~L. Jong, H.~Y. Lam, and C.~Y. Leow, ``Evaluation of {UAV} communication performance in urban environment: A study on air-to-ground propagation factors,'' in \emph{2025 IEEE 17th Malaysia International Conference on Communication (MICC)}, 2025, pp. 1--6.

\bibitem{rappaport}
T.~S. Rappaport, T.~E. Humphreys, and S.~Nie, ``Spectrum opportunities for the wireless future: From direct-to-device satellite applications to {6G} cellular,'' \emph{npj Wireless Technology}, vol.~1, no.~1, p.~8, 2025.

\bibitem{YastrebovaCastillo2024EASN}
A.~Yastrebova-Castillo, S.~Tocklin, H.~Kokkinen, M.~A. Ullah, M.~H\"oyhty\"a, and M.~Majanen, ``Integrated {UAS}-satellite communications in {6G}: An overview,'' in \emph{Proceedings of the 15th EASN International Conference}, 2024, accepted for publication.

\bibitem{3gppTR38101-5}
\BIBentryALTinterwordspacing
{3rd Generation Partnership Project (3GPP)}, ``{NR; User Equipment (UE) radio transmission and reception; Part 5: Satellite access Radio Frequency (RF) and performance requirements},'' 3GPP, Technical Report 38.101, 2022, release 17. [Online]. Available: \url{https://www.3gpp.org/ftp/Specs/archive/38_series/38.101-5}
\BIBentrySTDinterwordspacing

\bibitem{StarlinkWeather}
M.~Asad~Ullah, A.~Heikkinen, M.~Uitto, A.~Anttonen, and K.~Mikhaylov, ``Impact of weather on satellite communication: Evaluating {Starlink} resilience,'' in \emph{2025 IEEE 101st Vehicular Technology Conference (VTC2025-Spring)}, 2025, pp. 1--7.

\bibitem{ITU_R_P1410_2023}
\BIBentryALTinterwordspacing
{International Telecommunication Union (ITU)}, ``{Recommendation ITU-R P.1410-6: Propagation data and prediction methods required for the design of terrestrial broadband radio access systems operating in a frequency range from 3 GHz to 60 GHz},'' ITU-R, Recommendation P.1410-6, Aug. 2023. [Online]. Available: \url{https://www.itu.int/dms_pubrec/itu-r/rec/p/R-REC-P.1410-6-202308-I!!PDF-E.pdf}
\BIBentrySTDinterwordspacing

\bibitem{Akram}
A.~Al-Hourani, S.~Kandeepan, and S.~Lardner, ``Optimal {LAP} altitude for maximum coverage,'' \emph{IEEE Wireless Communications Letters}, vol.~3, no.~6, pp. 569--572, 2014.

\bibitem{Friis}
H.~Friis, ``A note on a simple transmission formula,'' \emph{Proceedings of the IRE}, vol.~34, no.~5, pp. 254--256, 1946.

\bibitem{3gppTS38521_1}
\BIBentryALTinterwordspacing
{3rd Generation Partnership Project (3GPP)}, ``{NR; User Equipment (UE) conformance specification; Radio transmission and reception; Part 1: Range 1 standalone},'' 3GPP, Technical Specification (TS) 38.521-1, 2022, version 17.5.0, Release 17, ETSI TS 138 521-1 V17.5.0, under change control. [Online]. Available: \url{https://portal.3gpp.org/desktopmodules/Specifications/SpecificationDetails.aspx?specificationId=3381}
\BIBentrySTDinterwordspacing

\bibitem{FCC_AST_Application}
{Federal Communications Commission (FCC)}, ``{Experimental License System Application (File No. 281537)},'' \url{https://apps.fcc.gov/els/GetAtt.html?id=281537&x=}, 2023, accessed: Feb. 2026.

\bibitem{SpaceX_TMobile_Technical_Narrative}
{SpaceX} and {T-Mobile}, ``{Technical Narrative: Direct-to-Cellular Service},'' \url{https://cdn.arstechnica.net/wp-content/uploads/2023/05/SpaceX-T-Mobile-Technical-Narrative.pdf}, 2023, accessed: Feb. 2026.

\bibitem{3gppTR38821}
\BIBentryALTinterwordspacing
{3rd Generation Partnership Project (3GPP)}, ``{Solutions for NR to Support Non-Terrestrial Networks (NTN)},'' 3GPP, Technical Report (TR) 38.821, 2020, release 16, under change control. [Online]. Available: \url{https://portal.3gpp.org/desktopmodules/Specifications/SpecificationDetails.aspx?specificationId=3525}
\BIBentrySTDinterwordspacing

\end{thebibliography}
\end{document}